\documentclass[12pt,showpacs,reprint,superscriptaddress,amsmath,amssymb]{iopart}
\usepackage{graphicx}

\usepackage{graphicx}
\usepackage{dcolumn}
\usepackage{bm}

\usepackage[dvips]{color}

\usepackage{iopams}  
\begin{document}

\title{Advanced Quantum Noise}

\author{Ulrich Vogl$^1$, Ryan T Glasser$^1$, Jeremy B Clark$^{1}$, Quentin Glorieux$^{1,2}$, Tian Li$^{1}$, Neil V  Corzo$^{1}$ and Paul D Lett$^{1}$}
\address{$^1$ Quantum Measurement Division, National Institute of Standards and Technology and
Joint Quantum Institute, NIST \& the University of Maryland, Gaithersburg, MD 20899 USA}
\address{$^{2}$Group of Applied Physics, University of Geneva, Chemin de Pinchat 22, CH-1211 Geneva, Switzerland}

\ead{ulrich.vogl@mpl.mpg.de}

\begin{abstract}
We use the quantum correlations of twin-beams of light to probe the added noise when one of the beams propagates through a medium with anomalous dispersion. The experiment is based on two successive four-wave mixing processes in rubidium vapor, which allow for the generation of bright two-mode-squeezed twin-beams followed by a controlled advancement while maintaining the shared quantum-correlations between the beams. The demonstrated effect allows the study of irreversible decoherence in a medium exhibiting anomalous dispersion, and for the first time shows the advancement of a bright nonclassical state of light.  The advancement and corresponding degradation of the quantum correlations are found to be operating near the fundamental quantum limit imposed by using a phase-insensitive amplifier.

\end{abstract}


\maketitle

\section{Introduction}

Anomalous dispersion in dielectric media can lead to negative group velocities and superluminal propagation of classical optical pulses.
This could on first glance seem to lead to a conflict with information (or Einstein)  causality, but it is actually a natural consequence of the causal transfer function of atomic media \cite{Milonni2007}.
One physical explanation that is sometimes given for the speed limitation on causal information transfer is that fundamental quantum processes will inevitably add sufficient noise to any communication channel to prevent information transfer faster than the speed of light  \cite{Milonni2007,1998Aharonov,ChiaoNoise,Narum,Molotkov2010}.
Here we demonstrate the use of  a bright, continuous-variable two-mode state of light that exhibits quantum correlations below the classical limit to investigate the advancement of optical signals in fast light media.
Using the correlated fluctuations in twin-beams  generated by four-wave mixing (4WM) we study the degradation of the quantum correlations as a function of the group velocity advancement due to anomalous dispersion.
We show that noise measurements on our experimental system  behave in quantitative agreement with the theoretical predictions based on a quantum--limited phase--insensitive optical amplifier.

Slow light properties (or positive group indices) produce temporal delays that can be investigated using quantum-correlated twin beams as well \cite{lukin2000,zac2001,van2003atomic,PhysRevLett.100.093601,2008PhRvL.100i3602A,marino2009tunable}.
These delays are produced while also adding noise, but without the same potential for controversy.
Such slow light media make it possible to delay optical pulses with a frequency bandwidth matched to the bandwidth of the normal dispersion by many pulse lengths.
With the other sign of the dispersion, and fast light media, the same physics of altered group velocities is at play  \cite{garrett1970,neifeld2003,2000Peatross,2002Wynne}, while the pulse advance is fundamentally much more constrained and the fundamental role of the noise becomes important to study as well \cite{Milonni2007,1998Aharonov,ChiaoNoise,Narum,Molotkov2010}.

\section{Fast light and quantum noise}

While there have been many experiments investigating the advancement of classical pulses, there are relatively few experiments that address this issue for  a non-classical state of light.  While some investigations have involved single photon propagation in fast--light media  \cite{singlephoton1,Franson2008,precursor2011}, here we use a continuous-variable bright quantum state to probe a fast--light medium.  Continuous--variable quantum states eliminate the need for photon--counting or number--resolving detection, and allow for high--efficiency direct detection schemes to be used.
Due to the lack of experimental investigations in this regime, it is an open question whether or not quantum correlations such as entanglement or squeezing could be detected after propagation through a fast light medium.
One can imagine that, due to the noise added by the phase-insensitive gain during the propagation, the quantum state would be immediately degraded and the quantum correlations fully destroyed given any advancement.
On the other hand, it might be possible for some degree of quantum correlation to survive, particularly for advancements short compared to the correlation times involved.
In this paper we address this question by sending half of a bipartite continuous variable quantum state through a fast--light medium.
This signal consists of fluctuations on one mode of a bright twin--beam state of light.
This mode taken by itself exhibits intensity fluctuations that are much larger than the standard quantum limit.
Due to the nature of the twin--beam state these random fluctuations are quantum correlated with the fluctuations in the other twin beam, which acts as a time--reference against which we  can calibrate the advancement.
The intensity cross-correlation between the beams will be shifted in time, depending on the relative advancement of the signal beam fluctuations and the survival of the quantum correlations  \cite{marino2009tunable}.

\begin{figure}[t!!!]
\includegraphics[width=\textwidth]{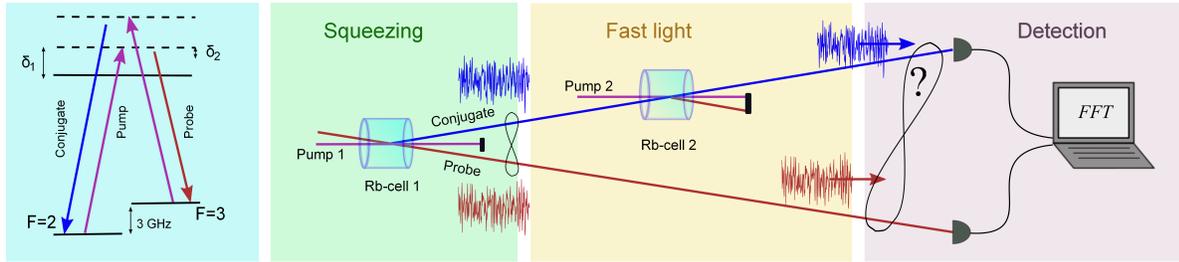}
\caption{\label{fig:epsart1} Experimental scheme. On the left we show the relevant transitions for the 4WM scheme with the one-photon detuning $\delta_1$ and the two-photon detuning $\delta_2$. Intensity-squeezed twin beams (probe and conjugate beams) are created via the 4WM scheme in the first Rb cell. The conjugate beam is fed into a second 4WM process in the second Rb cell with the second pump beam frequency slightly detuned from the first pump beam. The photocurrent from the two detectors is digitized and subsequently the intensity-difference spectrum is analysed via a discrete Fourier-transformation and the relative time lag between the beams is determined by their cross-correlation function.
}
\end{figure}

The semi-classical description of slow light and fast light propagation in atomic media is typically based on the expression for the refractive index of an atomic gas \cite{gauthier2007}, which takes the form
\begin{equation}
n(\omega)=1+\frac{g}{4\pi}\frac{\gamma}{\omega-\omega_0+\textit{i}\gamma},
\end{equation}
where $g$ is the gain coefficient, $\gamma$ the linewidth, $\omega_0$ is the center angular frequency of an optical transition, and $\omega=\frac{2\pi}{k}$ the angular frequency of the optical field ($k$ is the accompanying wave vector).  This results in the group index $n_g=n+\omega \frac{dn}{d\omega}$ and the corresponding group velocity $v_g=\frac{d\omega}{dk}=c/n_g$. In regions where the dispersion $\frac{d\omega}{dk}$ can be linearly expanded over a given bandwidth, optical pulses with matching frequency bandwidth can propagate nearly distortion-free while being delayed or advanced relative to a pulse traveling through vacuum.  This is quantified by the pulse peak delay
\begin{equation}
 \Delta T=\frac{L}{c}(n_g-1),
\end{equation}
 where $L$ is the propagation length and $c$ is the vacuum speed of light.  There is no known physical limit for the possible delay of optical pulses (i.e. for $n_g > 1$). The situation is different for advancing optical pulses (corresponding to $n_g < 1$), where a large relative advancement could seem to be in conflict with the relativistic causality principle. The seeming conflict was quickly identified and resolved by Sommerfeld and Brillouin  \cite{1914Sommerfeld, 1914Brillouin}, who showed that a superluminal group velocity in dispersive media cannot be used for any information transfer faster than $c$, and that the group velocity cannot be identified with the information velocity of light pulses, which is always strictly $c$.

 Triggered by advancements in experimental fast light systems, renewed discussions on the topic have appeared over time  \cite{garrett1970,neifeld2003,2000Peatross,2002Wynne,1982Chu,chiao1994,boyd2003,precursor2011,  1998Mitchell,dogariu2000,PhysRevA.73.033806}.
This led to new proposals to investigate the physical processes that actually work to prevent large pulse advances and superluminal signaling \cite{Milonni2007,1998Aharonov,ChiaoNoise,Narum,Molotkov2010}.
The usual discussion of (noiseless) classical signals being passed through a fast light medium involves the discussion of analytic signals.  While the peak of a pulse form may be advanced, it is said that any measurement of the faintest leading edge of the pulse, an analytic signal, will determine the following waveform.  Thus, the information was transmitted well before the peak, or any other obvious part of the waveform.  This, of course, implies that new information can only correspond to non-analytic points in the waveform, which also corresponds to an infinite frequency bandwidth to accommodate any discontinuities in the waveform or its derivatives, which can never fit into the finite linear dispersion regions that we can create.  While this tightly constrains the problems that can be discussed, the fact is that the tiniest of leading edges of a pulse waveform is never well-defined in practice, and one cannot create analytic optical signals.  At some point the signal is so small that noise dominates.  Equivalently, the  wave function of the optical pulse will contain less than a single photon, and the wave function can only be measured in a statistical sense.  In this case quantum field noise as well as detector noise becomes fundamental to even defining the problem.  In these scenarios the ability to discern signals is limited by the signal--to--noise ratio.  States of light that correspond to classical distributions of fields and their fluctuations result in a noise floor referred to as the standard quantum limit, which is a limiting factor in how precisely one can make measurements.  Coherent states of light fall into this category.
The use of quantum states of light can allow measurements to go beyond this limit by lowering the noise floor.
The bright twin-beam state of light used here exhibits this property, and results in a 2.5~dB reduction of noise below the shot noise limit \cite{footno}.

\section{Experimental scheme}
 
Our experimental scheme is shown in Fig.\,\ref{fig:epsart1}.
 Our system is based on 4WM in rubidium vapor, which is used to generate a pair of bright, strongly intensity-correlated twin beams.
 Each beam itself exhibits random intensity fluctuations, but the noise of the intensity-difference signal of the two photocurrents can be well below the standard quantum limit over a range of detection frequencies of a few MHz.
The intensity--difference noise of the twin beams can be expressed using the mean photon numbers of the probe and conjugate modes,$\langle \hat{n}_p\rangle$ and $\langle \hat{n}_c\rangle$  \cite{walls1986,Lukin2000b}:
\begin{equation}
%\sout{\langle\Delta(\hat{n}_a-\hat{n}_b)^2\rangle=\langle \Delta\hat{n}_a^2\rangle + \langle \Delta \hat{n}_b^2\rangle -2 \langle \hat{n}_a \hat{n}_b\rangle + 2 \langle \hat{n}_a\rangle\langle \hat{n}_b\rangle.}
%\\
\langle\Delta(\hat{n}_p-\hat{n}_c)^2\rangle=\langle \Delta\hat{n}_p^2\rangle + \langle \Delta \hat{n}_c^2\rangle -2 \langle \hat{n}_p \hat{n}_c\rangle + 2 \langle \hat{n}_p\rangle\langle \hat{n}_c\rangle.
\end{equation}
This variance of the two-mode squeezed state can be smaller than the variance obtained with a pair of coherent state beams with equal intensity, which in this context defines the standard quantum limit.

One of the beams is sent through a second 4WM process such that it experiences a region of anomalous dispersion resulting from a nearby gain line, which has a spectral overlap with the typical range where intensity-difference squeezing between the twin-beams is present.
Thus we can expect two things to occur:

i) Random intensity fluctuations on the beam that fit into the spectral bandwidth of the anomalous dispersion may be advanced according to Eq.(2), relative to their respective counterpart (the other beam of the twin-beam pair).

ii) The second 4WM process will inevitably introduce extra noise, which will either partially diminish or completely destroy the correlations between the twin beams.

The extra noise will consist partially of technical noise, such as scattered pump light, which can be minimized, and fundamental quantum noise.
The 4WM process used to generate fast light is phase-insensitive, and as such we can quantify this added fundamental noise by using the input-output relations for an ideal phase-insensitive amplifier \cite{jeffers1993quantum}.
 The expectation value of the output photon number $\langle \hat{n}_{out} \rangle $ is thereby connected to the input photon number $\langle \hat{n}_{in} \rangle $ and the gain $G$ by
\begin{equation}
\langle \hat{n}_{out} \rangle=G \langle \hat{n}_{in} \rangle +G -1.
\end{equation}
The variance on the output photon number is
\begin{equation}
\langle \Delta \hat{n}_{out}^2 \rangle = G^2 \langle \Delta \hat{n}_{in}^2 \rangle + G(G-1)(\langle \hat{n}_{in} \rangle +1).
\end{equation}
With the expressions Eqs.\,(3-5) the expected variance of the twin beams can be derived for the situation of one beam experiencing independent ideal phase-insensitive gain, which provides a lower bound for the added noise in this type of system.  These expressions, along with the refractive index describing a Lorentzian gain line (Eq.(1)), model the measured gain line and noise very well quantitatively, as seen in Fig.~2.  This indicates that the noise added to the advanced beam due to the nearby gain line is very close to the fundamental limit for an ideal phase-insensitive amplifier.

\begin{figure} [t!!]
\begin{center}
\includegraphics[width=12cm]{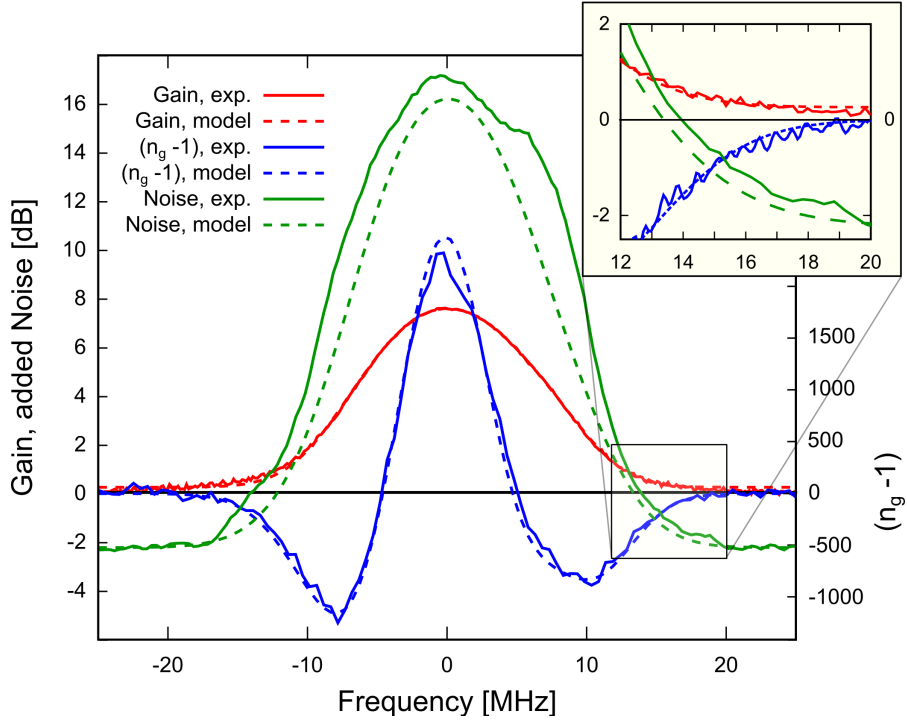}
\caption{\label{fig:epsart2} The red line represents a scan over the gain line in the fast light cell versus the detuning of the fast light pump laser (0\,dB is equal to a gain of 1). We simultaneously recorded the intensity-difference squeezing via balanced detection at a detection frequency of 750\,kHz (green line). The input squeezing starts from the baseline of intensity difference squeezing  -2.5\,dB below the standard quantum limit, which is marked by 0\,dB. The blue line shows $n_g-1$ derived from the measured gain profile. In the inset we show the region of interest for positive pump-detuning, where the intensity-difference noise is below the standard quantum limit and simultaneously the group index is less than zero. The model uses a Lorentzian fit to the observed gain line profile, the group index model follows from Eq.(1), the model for the added noise follows again from the observed gain, directly applied to Eqs.(3-5).
}
\end{center}
\end{figure}

The first stage of the experiment generates a pair of intensity-difference squeezed beams via a 4WM process in $^{85}$Rb vapor \cite{lett2007,pra2010,Glasser2012,Vogl2012,Glasser2012b,NJP2012,Vogl2012b}. The light beams involved are derived from a tapered amplifier system that is seeded with an external-cavity diode laser operating at 795\,nm.  A strong pump beam (200\,mW) is sent through a vapor cell (heated to 114\,$^{\circ}$C), and a weak probe beam (20\,$\mu$W), detuned by $\approx$ +3\,GHz, is injected at a small angle, as shown in Fig.\,\ref{fig:epsart1}.
The 4WM gain amplifies the probe beam while traveling through the rubidium vapor and generates a conjugate beam at $\approx$ -3\,GHz relative to the pump and in a separate spatial mode.
The coupled gain in the probe and conjugate modes results in two strongly intensity-correlated beams, which can be verified by direct intensity-difference detection with a pair of balanced photodiodes.

In the second step of the experiment one of the correlated beams is injected into a second 4WM process where the pump frequency is detuned a few MHz relative to the pump of the first 4WM process, resulting in additional gain and accompanying dispersion on the injected beam. The frequency of the second pump beam can be detuned independently of the frequency of the first pump beam. A full scan of the resulting gain line is shown in Fig.\,\ref{fig:epsart2}, with a peak gain of 7.5\,dB.
The resulting gain line is of approximate Lorentzian shape and has a typical full width at half maximum of 10\,MHz.

\section{Results}

The effect on the conjugate beam caused by traveling through this gain region can be described by a complex refractive index. 
The imaginary part describes the gain, which in turn adds noise proportional to the gain. 
This can be directly seen in the measured intensity-difference noise trace in Fig.\,\ref{fig:epsart2}.
Here the gain is plotted in red.
The theoretically predicted noise (dashed green) associated with phase--insensitive amplification is closely followed by the measured noise (solid green).  
It is known that nonlinear optical susceptibilities do not always satisfy Kramers--Kronig relations applicable to linear dielectric media.
For a nonlinear medium driven by a constant pump at a separate frequency from a weak probe, however, a related dispersion relation can be constructed.
In this case, as long as probe saturation or self--action effects are not important, a dispersion relation can be written.
The resulting theoretical fits for the frequency--dependent noise and index of refraction agree well with the measured data, as seen in Fig.~2.
The real part of the refractive index describes the modified dispersion and the altered group index, shown in blue.
The group index shows large positive values (slow light) near the center of the gain line, and negative values at the wings (fast light).
The standard equations describing a Lorentzian gain line (Eq.\,1) and noise added by phase-insensitive amplification (Eqs\,3-5) result in a good fit to the experimental data.

This characterization defines the detuning range for the pump used in the second 4WM process, relative to the first 4WM pump frequency.
We can expect the best chance for quantum correlations to survive is when the additional gain is close to 1 and only a small amount of noise is added.
This can be achieved in a region where the group index is negative on both wings of the gain line, at detunings ranging from -20\,MHz to -15\,MHz and from 14.5\,MHz to 20\,MHz. The gain in these regions varies from 1 to 1.25, with an experimental uncertainty of 0.05.  This sets the center frequency and bandwidth for which we can expect group advancement of a transmitted signal.
Frequency components of the signal that lie outside of this frequency bandwidth experience a delay instead of an advancement.

 A pair of amplified balanced detectors (detection efficiency 95\,\%) allows us to directly monitor the amount of intensity difference squeezing. Additionally, by acquiring the individual time traces of the photocurrents generated by the twin-beams we can measure the correlation between the two beams, giving us information about the relative advancement of the beam that passes through the fast light medium.
All detectors are at equal distances from the first 4WM process to within a few centimeters, and thus the distances are not a limiting factor in determining the delay or advancement.
 The signals are recorded simultaneously with a high-speed digital oscilloscope (1 million points per trace, 2.5\, giga--samples per second).
We take a discrete Fourier transform of the difference signal to derive the spectral noise power of the intensity-difference output signal.
 The shot noise level is determined by sending a pair of coherent states with the same total optical power as the generated twin beams into the detection path.
 Our setup allows for intensity--difference squeezing of -2.5\,dB below the shot noise limit, mainly limited by the available pump power and scattered pump light.

\begin{figure}
\begin{center}
\includegraphics[width=10cm]{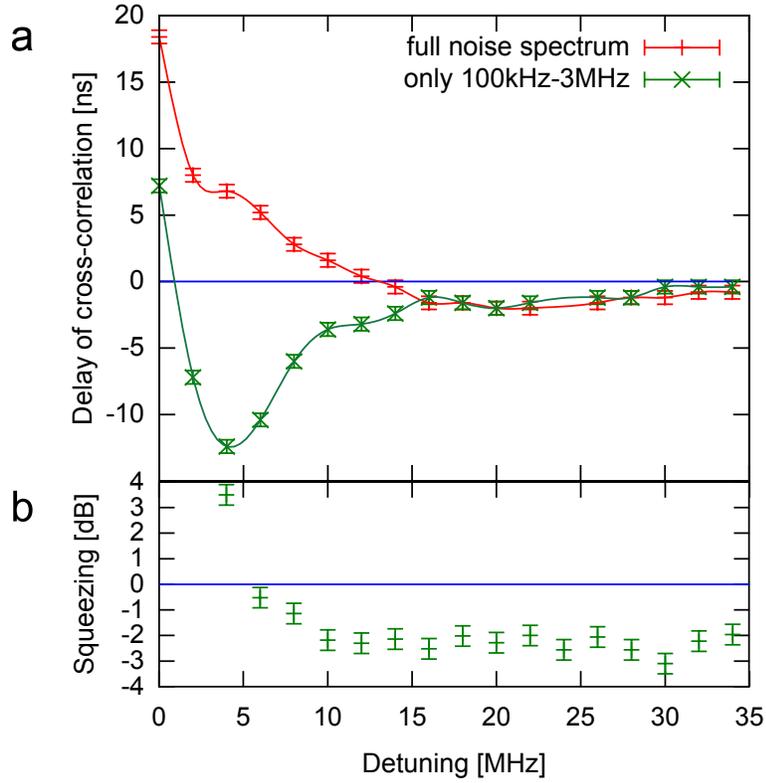}
\caption{\label{fig:epsart5}
 (a) Observed delay of the cross-correlation function versus the detuning of the pump beam of the fast-light process relative to the pump beam that generates the squeezed beams.
Pump powers are 200~mW and 300~mW respectively.  The pump diameters in the center of the first and second cells are $\approx$\,1\,mm and $\approx$\,1.5\,mm.
 The red points show the delay obtained by using the full spectrum of the correlated photon pairs ($\approx$ 20\,MHz).
 The green points show the cross--correlation for the noise bandwidth between 100\,kHz and 3\,MHz.
 (b) Simultaneously observed relative intensity squeezing in the bandwidth between 100\,kHz and 3\,MHz.
 }
 \end{center}
\end{figure}

\begin{figure}[ttt]
\begin{center}
\includegraphics[width=10.0cm]{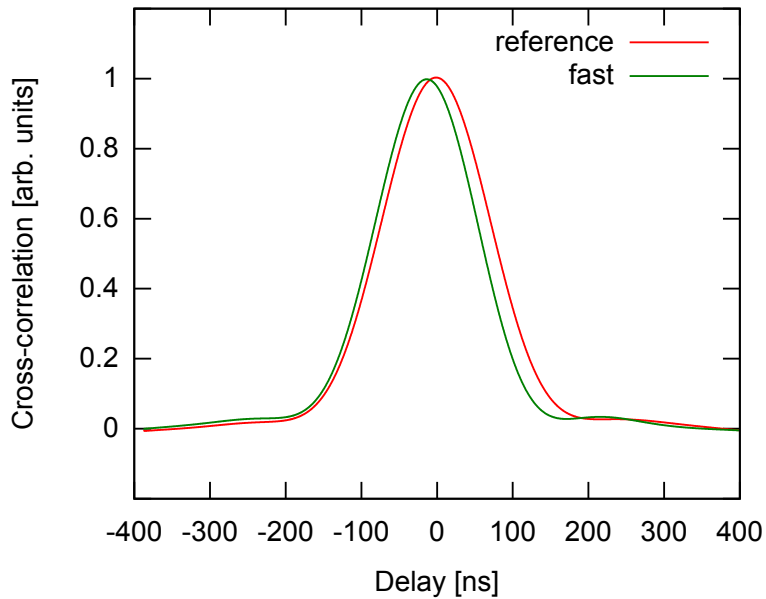}
\caption{\label{fig:epsart4} Cross-correlation between probe and conjugate beam for a relative detuning of the two pump beams of 6\,MHz.
 The cross-correlation shown is composed from the frequency spectrum from 100\,kHz to 3\,MHz, where we also observe relative intensity squeezing.
The correlations are advanced by 12\,ns.
}
\end{center}
\end{figure}

From the two individual time traces $i_1$ and $i_2$ we calculate the intensity cross--correlation function \cite{mandel1965coherence,PhysRevA.36.192}
\begin{equation}
C_{12}(t)=\int i_1(\tau)i_2(t+\tau)\textit{d}\tau
\end{equation}
with a resolution of 400\,ps, where $i_1$, $i_2$ denote the individual photocurrents. We use the cross-correlation of the advanced beam with the reference twin beam as the signature for the presence of advancement or delay, which is given by
\begin{equation}
\Delta t = t|_{{\max(C_{12}^{{Fast}})}}-t|_{{\max(C_{12}^{{Ref}})}}.
\end{equation}

To verify if squeezing between the two beams can be maintained when one of the beams is advanced, we now set the detuning of the fast-light pump in the region where we can expect (from the line scan of Fig.\,\ref{fig:epsart2}) to have both significant anomalous dispersion and an extra gain not far above 1.
In addition to the intrinsic added noise due to the phase insensitive gain, scattered pump light contributes an additional $\approx$~0.2\,dB of extra noise.

A sampling over a wide range of detunings of the fast-light pump, where the measured gain on the conjugate beam is less than 1.1 (0.41\,dB), is shown in Fig.\,\ref{fig:epsart5}(a).
Shown is the measured delay according to Eq.\,7 versus the measured squeezing integrated over the bandwidth between 100\,kHz to 3\,MHz.
We obtain a relative advancement of up to 12\,ns while still maintaining some relative intensity squeezing of the twin beams (see Fig.\,\ref{fig:epsart5}(b)).
 The measured data is quantitatively consistent with the additional noise being the minimum amount that must be added by a quantum--limited phase-insensitive amplifier  (see Eq.(5)).
Following Eq.(2), the advancement of 12\,ns corresponds to a group index on the order of -150, which is in good quantitative agreement with the values shown in Fig.\,2.
 Also the characteristic shape of the group index versus the two-photon detuning shown in Fig.\, 3 agrees well with a Lorentzian gain line model (see Eq.(1)).

A typical example for the resulting intensity cross-correlation function is shown in
Fig.\,\ref{fig:epsart4}.
The red line shows the cross-correlation for the undisturbed input beam pair with a relative intensity squeezing of -2.5\,dB.
 The green line shows an example for a cross-correlation between the beams, when one beam is passed near-resonantly (i.e. for a detuning of 6\,MHz in Fig.~3) through the fast-light region of the second 4WM process.
We compute the cross--correlation function only using frequencies in the spectral region where we observe quantum correlations by applying a digital filter to the recorded data .
 The same filtering is applied to all of the data, in both the reference and fast--light cases.
We use a Hanning window with the 3\,dB roll-off points at 100\,kHz and 3\,MHz before the cross-correlation is computed.
As can be seen in Fig.\,\ref{fig:epsart5}(a), in this case the sign of the advancement/delay of the cross-correlation changes for different spectral bands of the correlation.

We interpret this behavior as follows: The first 4WM process creates correlated photon pairs over a bandwidth of $\approx$\,20\,MHz, which is determined by the spectral width of the gain lines for the probe and the conjugate of the 4WM process. The second 4WM process operates at a center frequency that is detuned a few MHz from the first process. Consequently different parts of the spectrum will experience normal dispersion or anomalous dispersion when passing through the second cell.
We can confirm this interpretation by filtering the photocurrents to the band from 100\,kHz to 3\,MHz, as shown in Fig.~(3)a.
The twin-beams exhibit squeezing over this entire bandwidth, with a resulting cross--correlation FWHM of $\approx$\,165\,ns.
This spectral band also experiences only anomalous dispersion for detunings of the second pump greater than 3\,MHz, and the resulting cross-correlation is advanced.
  For detunings of the fast-light pump greater than 15\,MHz the advancement of the cross-correlation is not much affected by the choice of bandwidth, as here the full bandwidth of the photon pairs generated in the first 4WM process fits into the region of anomalous dispersion of the second 4WM process.
For zero detuning we reproduce an experimental situation similar to the one described in \cite{marino2009tunable}, where most of the correlations are delayed and strongly degraded by added noise.
 In this case the spectral region of the quantum correlations that exhibit slow light are closer to the center of the gain line and therefore suffer from more excess noise.

\section{Conclusion}

In summary, we have performed an experiment in which we sent one arm of a two-mode squeezed state through an atomic medium with anomalous dispersion.
We have shown that the noise added due to the advancement using this system operates near the fundamental limit imposed by using a phase-insensitive amplifier as the fast--light medium.
  Additionally, the present experiment shows that it is, in principle, possible to advance one of the two modes in a bipartite quantum state and still maintain some degree of nonclassical correlation, albeit for advances short compared to the average correlation time involved.
The effect demonstrated here can lead to a better understanding of the physical processes that contribute to the ``peaceful coexistence between quantum mechanics and relativity" \cite{Shimony}.
An extension of this work would be to investigate the dispersive and noise properties of a phase-sensitive amplifier \cite{2012Corzo}.

\section{Acknowledgments}
This work was supported by the Air Force Office of Scientific Research.
UV acknowledges support from the Alexander von Humboldt Foundation.
 This research was performed while RG held a National Research Council Research Associateship Award at NIST.
 QG is supported by the Marie Curie IOF FP7 Program - (Multimem - 300632).


\section*{References}
\begin{thebibliography}{10}



\bibitem{Milonni2007}
Boyd R W, Shi Z and Milonni P W 2010
{\it Journal of Optics} \textbf{12} 104007


\bibitem{1998Aharonov}
{Aharonov} Y, {Reznik} B and {Stern} A 1998
%{\it \emph{{Quantum Limitations on Superluminal Propagation}}.
{\it Phys. Rev. Lett.} \textbf{81} 2190 

\bibitem{ChiaoNoise}
{Kuzmich} A, {Dogariu} A,  {Wang} L J, {Milonni} P W and {Chiao} R Y 2001
%{\it \emph{{Signal Velocity, Causality, and Quantum Noise in Superluminal  Light Pulse Propagation}}.
{\it Phys. Rev. Lett.} \textbf{86} 3925 


\bibitem{Narum}
Boyd R W and Narum P 2001
%{\it \emph{Slow- and fast-light: fundamental limitations}.
{\it Journal of Modern Optics} \textbf{54} 2403 

\bibitem{Molotkov2010}
Molotkov S 2010
%{\it \emph{On the supraluminal group velocity and the transmission of  information}.
{\it JETP Letter} \textbf{91} 693 
%{\it 10.1134/S0021364010120155.

\bibitem{lukin2000}
Phillips D F, Fleischhauer A, Mair A, Walsworth R L and Lukin M D 2001
%{\it \emph{Storage of Light in Atomic Vapor}.
{\it Phys. Rev. Lett.} \textbf{86} 783

\bibitem{zac2001}
{Liu} C, {Dutton} Z, {Behroozi} C H and {Hau} L V 2001
%{\it \emph{{Observation of coherent optical information storage in an  atomic medium using halted light pulses}}.
{\it Nature} \textbf{409} 490 

\bibitem{van2003atomic}
van~der Wal C, Eisaman M, Andr{\'e} A, Walsworth R, Phillips D, Zibrov A
  and Lukin M 2003
%{\it \emph{Atomic memory for correlated photon states}.
{\it Science} \textbf{301} 196 

\bibitem{PhysRevLett.100.093601}
Honda K, Akamatsu D, Arikawa M, Yokoi Y, Akiba K, Nagatsuka S,
  Tanimura T, Furusawa A and Kozuma M 2008
%{\it \emph{Storage and Retrieval of a Squeezed Vacuum}.
{\it Phys. Rev. Lett.} \textbf{100} 093601

\bibitem{2008PhRvL.100i3602A}
{Appel} J, {Figueroa} E, {Korystov} D, {Lobino} M and {Lvovsky} A I 2008
%{\it \emph{{Quantum Memory for Squeezed Light}}.
{\it Phys. Rev. Lett.} \textbf{100} 093602 

\bibitem{marino2009tunable}
Marino A, Pooser R, Boyer V and Lett P 2009
%{\it \emph{Tunable delay of Einstein--Podolsky--Rosen entanglement}.
{\it Nature} \textbf{457} 859 

\bibitem{garrett1970}
Garrett C G B and McCumber D E 1970
%{\it \emph{Propagation of a Gaussian Light Pulse through an Anomalous  Dispersion Medium}.
{\it Phys. Rev. A} \textbf{1} 305 

\bibitem{neifeld2003}
{Stenner} M D,{Gauthier} D J and {Neifeld} M A 2003
%{\it \emph{{The speed of information in a `fast-light' optical medium}}.
{\it Nature} \textbf{425} 695 

\bibitem{2000Peatross}
{Peatross} J, {Glasgow} S A and {Ware} M 2000
%{\it \emph{{Average Energy Flow of Optical Pulses in Dispersive Media}}.
{\it Phys. Rev. Lett.} \textbf{84} 2370 

\bibitem{2002Wynne}
{Wynne} K 2002
%{\it \emph{{Causality and the nature of information}}.
{\it Optics Communications} \textbf{209} 85 

\bibitem{singlephoton1}
{Steinberg} A M, {Kwiat} P G and {Chiao} R Y 1993
{\it Phys. Rev. Lett.} \textbf{71} 708 

\bibitem{Franson2008}
Franson J D 2008
{\it J. Mod. Opt.} \textbf{55} 2117 

\bibitem{precursor2011}
Zhang S, Che J F, Liu C, Loy M M T, Wong G K L and Du S 2011
{\it Phys. Rev. Lett} \textbf{106} 243602 

\bibitem{gauthier2007}
Boyd R W and Gauthier D J 
{\it Chapter 6: Slow and fast light, Vol~43 of Progress in
  Optics.} (Elsevier, 2002)

\bibitem{1914Sommerfeld}
{Sommerfeld} A 1914
%{\it \emph{{{\"U}ber die Fortpflanzung des Lichtes in dispergierenden  Medien}}.
{\it Annalen der Physik} \textbf{349} 177 

\bibitem{1914Brillouin}
{Brillouin} L 1914
%{\it \emph{{{\"U}ber die Fortpflanzung des Lichtes in dispergierenden  Medien}}.
{\it Annalen der Physik} \textbf{349} 203 

\bibitem{1982Chu}
{Chu} S and {Wong} S 1982
%{\it \emph{{Linear Pulse Propagation in an Absorbing Medium}}.
{\it Phys. Rev. Lett.} \textbf{48} 738 

\bibitem{chiao1994}
Steinberg A M and Chiao R Y 1994
%{\it \emph{Dispersionless, highly superluminal propagation in a medium  with a gain doublet}.
{\it Phys. Rev. A} \textbf{49} 2071 

\bibitem{boyd2003}
Bigelow M S, Lepeshkin N N and Boyd R W 2003
%{\it \emph{Superluminal and Slow Light Propagation in a Room-Temperature  Solid}.
{\it Science} \textbf{301} 200 

\bibitem{1998Mitchell}
{Garrison} J C, {Mitchell} M W,  {Chiao} R Y and {Bolda} E L 1998
%{\it \emph{{Superluminal signals: causal loop paradoxes revisited}}.
{\it Physics Letters A} \textbf{245} 19 

\bibitem{dogariu2000}
{Wang} L J, {Kuzmich} A and {Dogariu} A 2000
%{\it \emph{{Gain-assisted superluminal light propagation}}.
{\it Nature} \textbf{406} 277 

\bibitem{PhysRevA.73.033806}
Lezama A, Akulshin A M, Sidorov A I and Hannaford P 2006
%{\it \emph{Storage and retrieval of light pulses in atomic media with  ``slow'' and ``fast'' light}.
{\it Phys. Rev. A} \textbf{73} 033806 

\bibitem{footno}
We remind the reader that the entanglement property of a bipartite state does not allow superluminal signaling [c.f. \textit{J.~S. Bell., Speakable and Unspeakable in Quantum Mechanics: Collected  papers on quantum philosophy} (Cambridge University Press, 2004)], and this is not changed by including a delay or an advance in one beam path.



%\bibitem{zurek1982single}
%W.~Zurek and W.~Wooters.
%%{\it \emph{A single quantum cannot be cloned}.
%{\it Nature} \textbf{299}, 802 (1982).

%\bibitem{bell2004speakable}
%J.~S. Bell.
%{\it \emph{Speakable and Unspeakable in Quantum Mechanics: Collected  papers on quantum philosophy} (Cambridge University Press, 2004).

\bibitem{walls1986}
Reid M D and Walls D F 1986
%{\it \emph{Quantum theory of nondegenerate four-wave mixing}.
{\it Phys. Rev. A} \textbf{34} 4929 
%
\bibitem{Lukin2000b}
Lukin M, Hemmer P and Scully M
{\it Advances In Atomic, Molecular, and Optical Physics, Vol 42. (Academic Press,
  2000)}
%
\bibitem{jeffers1993quantum}
Jeffers J, Imoto N and Loudon R 1993
%{\it \emph{Quantum optics of traveling-wave attenuators and amplifiers}.
{\it Phys. Rev. A} \textbf{47} 3346 

\bibitem{lett2007}
McCormick C F, Boyer V, Arimondo E and Lett P D 2007
%{\it \emph{Strong relative intensity squeezing by four-wave mixing in  rubidium vapor}.
{\it Opt. Lett.} \textbf{32} 178 


\bibitem{pra2010}
Glorieux Q, Dubessy R, Guibal S, Guidoni L, Likforman J-P, Coudreau T 2010
{\it Phys. Rev. A} \textbf{82} 033819 

\bibitem{Glasser2012}
Glasser R T, Vogl U and Lett P D 2012
%{\it \emph{Stimulated Generation of Superluminal Light Pulses via  Four-Wave Mixing}.
{\it Phys. Rev. Lett.} \textbf{108} 173902 

\bibitem{Vogl2012}
Vogl U, Glasser R T and Lett P D 2012
%{\it \emph{Advanced detection of information in optical pulses with negative group velocity}.
{\it Phys. Rev. A} \textbf{86} 031806 

\bibitem{Glasser2012b}
Glasser R T, Vogl U and Lett P D 2012
%{\it \emph{{Demonstration of images with negative group velocities}}.
{\it Optics Express} \textbf{20} 13702 

\bibitem{NJP2012}
Glorieux Q, Clark J B, Corzo-Trejo N, and Lett P D 2012
{\it New Journal of Physics} \textbf{14} 123024 

\bibitem{Vogl2012b}
Vogl U, Glasser R T, Glorieux Q, Clark J B, Corzo N V and Lett P D 2013
%{\it \emph{Experimental characterization of Gaussian quantum discord  generated by four-wave mixing}.
{\it Phys. Rev. A} \textbf{87} 010101 


\bibitem{mandel1965coherence}
Mandel L and Wolf E 1965
%{\it \emph{Coherence properties of optical fields}.
{\it Reviews of Modern Physics} \textbf{37} 231 

\bibitem{PhysRevA.36.192}
Ou Z Y, Hong C K and Mandel L 1987
%{\it \emph{Detection of squeezed states by cross correlation}.
{\it Phys. Rev. A} \textbf{36} 192 

\bibitem{Shimony}
Shimony A
{\it in: S. Kamefuchi (Ed.), Foundations of Quantum Mechanics in the Light of New Technology, Phys. Soc. Japan, Tokyo, 1983}

%
%
%\bibitem{adesso2010quantum}
%G.~Adesso and A.~Datta.
%%{\it \emph{Quantum versus classical correlations in Gaussian states}.
%{\it Phys. Rev. Lett.} \textbf{105}, 30501 (2010).
%

%
%\bibitem{Deutsch91}
%D.~Deutsch.
%%{\it \emph{Quantum mechanics near closed timelike lines}.
%{\it Phys. Rev. D} \textbf{44}, 3197 (1991).
%
%\bibitem{2012NatCo...3E1092O}
%O.~{Oreshkov}, F.~{Costa} and {\v C}.~{Brukner}.
%%{\it \emph{{Quantum correlations with no causal order}}.
%{\it Nature Communications} \textbf{3}, 1092 (2012).
%
%\bibitem{bennett2009can}
%C.~H. Bennett, D.~Leung, G.~Smith and J.~A. Smolin.
%%{\it \emph{Can closed timelike curves or nonlinear quantum mechanics  improve quantum state discrimination or help solve hard problems?}
%{\it Phys. Rev. Lett.} \textbf{103}, 170502 (2009).
%
%\bibitem{PhysRevLett.110.060501}
%J.~L. Pienaar, T.~C. Ralph and C.~R. Myers.
%%{\it \emph{Open Timelike Curves Violate Heisenberg's Uncertainty  Principle}.
%{\it Phys. Rev. Lett.} \textbf{110}, 060501 (2013).

\bibitem{2012Corzo}
Corzo N V, Marino A M, Jones K M and Lett P D 2012
%{\it \emph{Noiseless Optical Amplifier Operating on Hundreds of Spatial  Modes}.
{\it Phys. Rev. Lett.} \textbf{109} 043602 



\end{thebibliography}
\end{document}